\documentclass[prl,twocolumn,superscriptaddress,showpacs,preprintnumbers,amsmath,amssymb]{revtex4}
\usepackage{graphicx}% Include figure files
\usepackage{color}
\usepackage{natbib}
\begin{document}

\title{Quasi-Superradiant Soliton State of Matter in Quantum Metamaterials}
\author{H. Asai}
\affiliation{Department of Physics, Loughborough University, Leicestershire, LE11 3TU, United Kingdom}
\affiliation{Electronics and Photonics Research Institute (ESPRIT), National Institute of 
Advanced Industrial Science and Technology (AIST), Tsukuba, Ibaraki 305-8568, Japan}
\author{S. Kawabata}
\affiliation{Electronics and Photonics Research Institute (ESPRIT), National Institute of 
Advanced Industrial Science and Technology (AIST), Tsukuba, Ibaraki 305-8568, Japan}
\author{A.M. Zagoskin}
\affiliation{Department of Physics, Loughborough University, Leicestershire, LE11 3TU, United Kingdom}
\affiliation{Theoretical Physics and Quantum Technologies Department,
Moscow Institute for Steel and Alloys, 119049 Moscow, Russia}
\author{S.E. Savel'ev}
\affiliation{Department of Physics, Loughborough University, Leicestershire, LE11 3TU, United Kingdom}
\begin{abstract}

Strong interaction of a system of quantum emitters (e.g., two-level atoms) with electromagnetic field induces specific correlations in the system accompanied by a drastic insrease of emitted radiation (superradiation or superfluorescence). Despite the fact that since its prediction this phenomenon was subject to a  
vigorous experimental and theoretical research, there remain open question, in particular, concerning the possibility of a first order phase transition to the superradiant state from the vacuum state. In systems of natural and charge-based artificial atome this transition is prohibited by ``no-go" theorems. Here we demonstrate numerically a similar transition in a one-dimensional quantum metamaterial - a chain of artificial atoms (qubits) strongly interacting with classical electromagnetic fields in a transmission line. The system switches from vacuum state with zero classical electromagnetic fields and all qubits being in the ground state to the quasi-superradiant (QS) phase with one or several magnetic solitons and finite average occupation of qubit excited states along the transmission line. A quantum metamaterial in the QS phase circumvents the ``no-go" restrictions by considerably decreasing its total energy relative to the vacuum state by exciting nonlinear electromagnetic solitons with many nonlinearly coupled electromagnetic modes in the presence of external magnetic field. 
\end{abstract}
\maketitle

\begin{figure}[h]
%\vspace{-1cm}
\includegraphics[width=\linewidth]{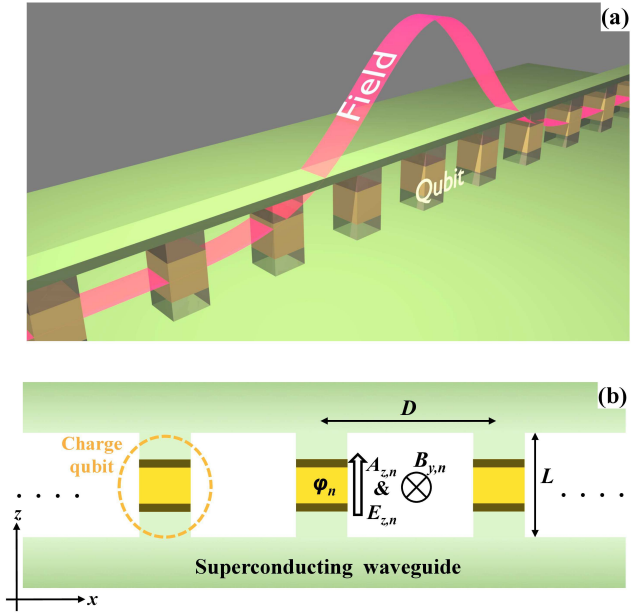}
\caption{\label{fig1}(color online) {\bf One-dimensional quantum metamaterial:} (a) A chain of superconducting charge qubits playing the role of artificial atoms in a Josephson transmission line formed between two superconducting banks connected via superconducting islands (charge qubits). Their quantum states are determined by the number of additional Cooper pairs present on an island. The (green) wave represent distribution of the electromagnetic field in the structure. 
(b) The schematic view of the structure. The magnetic field penetrates through the openings between the superconducting islands. The electric and magnetic fields and the vector potential are assumed constant within each opening. The additional gate electrodes required to control and tune the quantum state of the qubits are not shown. } 
\end{figure}

The interaction of light and matter in artificial optical media is the focus of a significant research effort (see, e.g., \cite{strongcoupl0,strongcoupl1,strongcoupl2,strongcoupl3,strongcoupl4, strongcoupl5}). The strong light-matter interaction in such systems make possible such effects as unusual photon collapse-and-revivals\cite{strongcoupl6}, Schr\"odinger-cat states\cite{franco}, non-classical radiation \cite{radiation}, unusual Casimir effect and pseudo-vacuum states\cite{psudovac}. For a subclass of these media with extended spatial quantum coherence (quantum metamaterials\cite{mi,Felbacq,science-lasing,macha,shapiro}) a number of novel phenomena are predicted, which do not have place in natural materials and classical metamaterials. This adds a new dimension to the long-standing discussion about the possibility of a superradiant transition in the system of atoms strongly interacting with electromagnetic waves in a cavity\cite{dicke,superrad,no-go1,no-go2,no-go3,nat-com,no-go-general}. In particular,  Ref.\cite{no-go-general} extends the "no-go" theorem to circuit QED with charge (but not flux) qubits.

\begin{figure}[h]
\includegraphics[width=7.7cm]{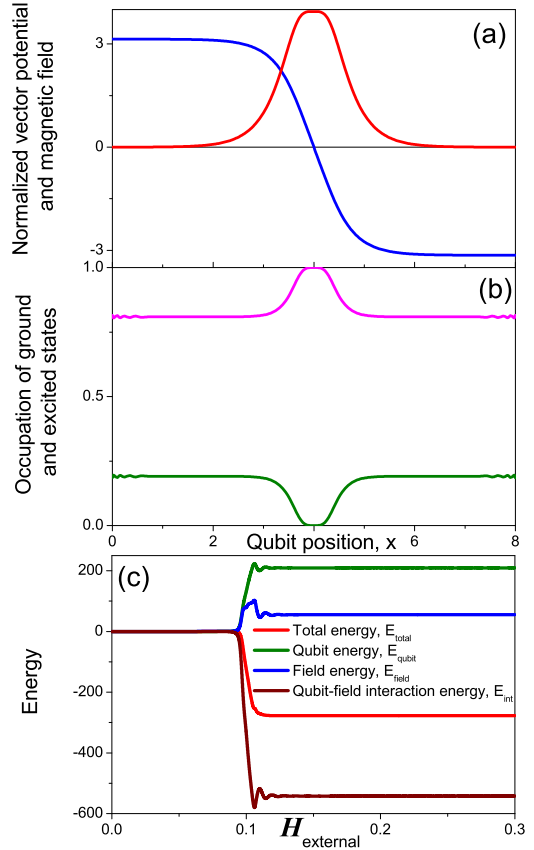}
\caption{\label{fig2}(color online) {\bf Quasi-superradiant phase:} (a) Distribution of dimensionless vector potential $a_{z,n}$ and magnetic field in the quantum Josephson transmission line, the variation of $a_{z,n}$
from $-\pi$ to $\pi$ indicates that the quasi-superradiant soliton carry one flux quantum. (b) Distribution of occupation probabilities for ground (bottom (green) curve) and excited (upper (magenta) curve) qubit states, the strong variation of the qubit state occupation occurs at the soliton center resulting in  pinning of the soliton. (c) Evolution of the qubit energy $E_{\rm qubit}$, the field energy $E_{\rm field}$, and the qubit-field interaction energy $E_{\rm int}$ as a function of the weak external magnetic field $H_{\rm ext}$. Even though the field
and qubit energies, $E_{\rm qubit}$ and $E_{\rm field}$, both increase at the transition, the total energy of the quasi-superradiant phase, $E_{\rm total}$, decreases due to a sharp drop of the interaction energy, $E_{\rm int}$.    
  Parameters used in simulations are: $s=1, \beta=0.25, \epsilon=\pi, l=L/\lambda=0.05, \gamma=0.25$ and the total number of qubits is 400.}
\end{figure}

\begin{figure}[h]
\includegraphics[width=8cm]{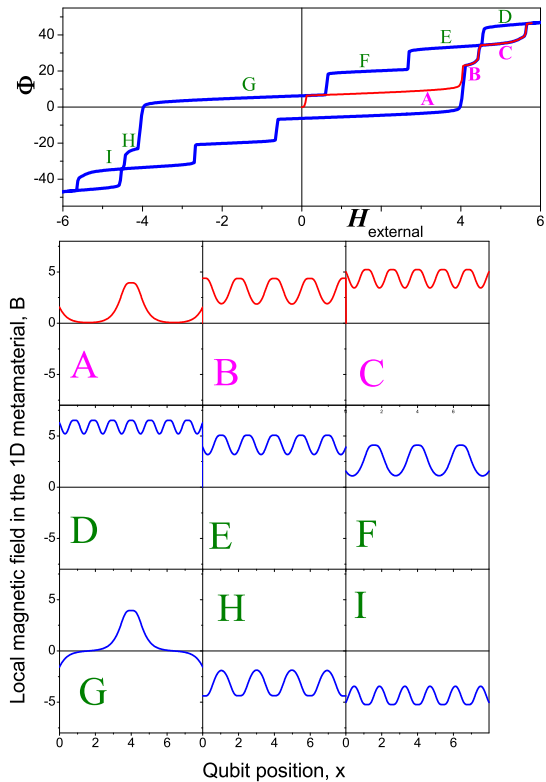}
\caption{\label{fig3}(color online){\bf Remagnetization loop as a sequence of the structural transitions:} (a) Evolution of the total flux $\Phi$ 
trapped in the  quantum transmission line as a function of the external magnetic field $H_{\rm ext}$. The (red) curve ABC starting at origin is the ``virgin curve'' which the system passes only once after its initialization in the vacuum state at zero external field, $H_{\rm ext}=0$. Applying a very weak external field results in the transition to the one-soliton phase (panel A) which then goes through a sequence of transitions to the phases 
with 3 (panel B), 5 (panel C) and 7 (panel D) solitons. On the returning branch the system passes through the states with 5 (panel E), 3 (panel F), 1 (panel G),
-3 (i.e., three anti-soliton) (panel H), -5 (five anti-solitons) (panel I) and -7 solitons. repeated cycling of $H_{\rm ext}$ forms the steady state (blue)
loop CD...I...B. Note that the state with one soliton or one anti-soliton is the ground state at $H_{\rm ext}=0$ with spontaneous symmetry breaking to either 1 or -1 soliton, while the vacuum (Meissner) state with no solitons is a metastable state. 
All parameters are the same as in Fig. 2, while the sweeping rate $dH_{\rm ext}/dt=2.5\times 10^{-4} 
\omega_J \Phi_0/\pi D\lambda$.}
\end{figure}

\begin{figure}[h]
\includegraphics[width=\linewidth]{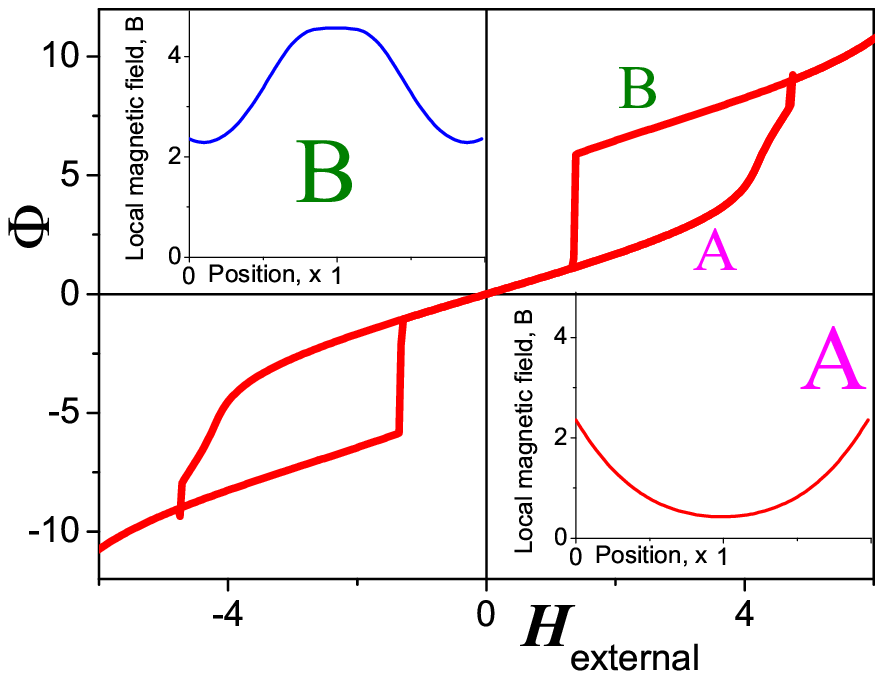}
\caption{\label{fig4}(color online){\bf ``Butterfly'' remagnetization curve:} $\Phi(H_{\rm ext})$ dependence for the same parameters as in Fig. 3, but for a shorter
transmission line (100 qubits). Note that there is no soliton phase at zero external magnetic field, $\Phi(H_{\rm ext}=0)=0$ (see inset A for the 
Meissner phase). Transition to the one-soliton phase (inset B) occurs at a relatively high magnetic field. On the return branch, the soliton
phase switches back to the Meissner phase before the external field $H_{\rm ext}$ drops to zero.}
\end{figure}

It was predicted\cite{dicke,superrad} that in a cavity containing many two-level atoms there exists a possibility of a transition from the vacuum state (with no photons  and all atoms being in the ground state) to the superradiant state (with nonzero photon occupation number accompanied by atom excitations) has been as the light-atom interaction increases. This topic remained subject of vigorous investigation ever since \cite{haroche,brandes}. Soon after this prediction it was pointed out  that the predicted transition is ``an interesting artefact'' \cite{no-go1}, caused by neglecting the term quadratic in the electromagnetic vector potential in the light-matter interaction Hamiltonian. Taking it into account eliminates the superradiant state.
This ``no--go'' theorem (see e.g., \cite{no-go1,no-go2,no-go3}) was claimed not to apply to both bosonic and fermionic artificial-atom arrays (in particular, in case of circuit QED), at least when driven by a laser to a non-equilibrium state \cite{nat-com,pump1,pump2,keeling-pump,keeling}. Nevertheless in \cite{no-go-general} the theorem was restated for charge-qubit based circuit-QED systems. The precise requirements  to a system capable of undergoing the superradiant phase transition\cite{nat-com,no-go-general} and possible connections between the superradiance and similar phenomena such as the dynamical Casimir effect \cite{Vacanti2012,berdiyorov} or the essentially non-classical spontaneous radiation \cite{radiation,psudovac} are still being vigorously debated, and the investigation of these and related phenomena in artificial structures is highly relevant. In particular, it was recently predicted that the electromagnetic pulse propagation in a one-dimensional superconducting quantum metamaterial could lead to lasing \cite{Asai} and superradiance and electromagnetic field-induced transparence effects\cite{Ivic}. 

A quantum metamaterial (QMM) is a {\it globally quantum coherent} array of artificial atoms with a limited control of their quantum state \cite{Felbacq}. In this article we consider superradiant-like transition in the presence of external magnetic field in such a structure. Though the basic properties of a QMM are qualitatively independent on its specific realization, here we use a model of a superconducting one-dimensional QMM essentially identical to the one of \cite{mi,Asai,Ivic}: a one-dimensional chain of charge qubits between two superconducting banks and interacting with electromagnetic fields, but not directly with each other (Fig. 1). In contrast to the earlier perturbative approach \cite{mi}, only applicable in the weak coupling limit, our numerical simulations allow  to consider the strong field-qubit coupling limit. As the result we find a phase transition from the vacuum state to the new  {\it quasi-superradiant phase}, which appears when the field-qubit coupling strength crosses  a certain threshold in the presence of an external magnetic field.
 
As the field-qubit coupling increases, the vacuum state with zero electromagnetic field and zero occupation of excited qubit states becomes unstable, and the system undergoes the first order phase transition to the state with magnetic solitons and a spatially varying occupation of the excited qubit states. The critical coupling strength can be reduced by increasing the external magnetic field $H_{\rm ext}$, which also generates structural transitions between states with different number of solitons. When cycling the external magnetic field around zero, the soliton number can be ether zero or nonzero at $H_{\rm ext}=0$ resulting in a reach variety of remagnetization loops associated with quantum pinning of solitons on the spatial variations of the qubit level occupation which, in turn, is generated by the solitons themselves. 

{\it Model---} The system in question is a quantum counterpart of the standard Josephson transmission line \cite{ustinov} where Josephson junctions linking two superconducting banks are replaced by charge qubits (Fig. 1), i.e., small superconducting islands connected to both long superconducting banks with high-resistance tunneling junctions ($R > R_Q \approx 12$ k$\Omega$), and with controlled potential bias with respect to the banks \cite{doudle}. The quantum states of an island differ by the number of extra Cooper pairs on them, and are coupled to the electromagnetic field through their electric charge. 

As in \cite{mi} we describe qubits quantum mechanically while treating the electromagnetic fields in the transmission line classically (in line with the standard semiclassical approach to atom-field interactions, valid in case of strong enough fields \cite{blokhintsev}). We direct the vector potential $A_z$ across the junctions (along $z$-axis) and assume it to be constant in each "cell" between two adjacent qubits. 

We start from the classical expression for the system's total (electromagnetic, electrostatic and Josephson) energy as a function of the phases $\phi_n$ of the superconducting order parameters on the islands:
$\mathcal{E}_{total} = \sum_n \mathcal{E}_n$ with 
\begin{eqnarray}
\mathcal{E}_n &=& \frac{C}{2} \frac{\hbar^2}{4e^2} \biggl[ \Bigl( \frac{d\phi_n}{dt} + \frac{\pi D}{\Phi_0} \frac{dA_{z,n}}{dt} \Bigr)^2 + \Bigl(\frac{d\phi_n}{dt} - \frac{\pi D}{\Phi_0} \frac{dA_{z,n}}{dt} \Bigr)^2 \biggr] \nonumber \\
&-& E_J \biggl[ \cos{\Bigl(\phi_n + \frac{\pi D A_{z,n}}{ \Phi_0} \Bigr)} + \cos{\Bigl(\phi_n - \frac{\pi D A_{z,n}}{\Phi_0} \Bigr)} \biggr] \nonumber \\
&+& \frac{8 \pi}{ D L} \biggl[ \frac{A_{z,n+1} - A_{z,n}}{L} \biggr]^2.
\label{total}
\end{eqnarray} 
Here, the index $n$ corresponds to the qubit number in the chain, $C$ is the junction capacitance, $\Phi_0 = hc/2e$ is the flux quantum, 
$E_J = \frac{I_c \Phi_0}{2 \pi c}$ is the Josephson coupling energy, $I_c$ is the critical current of the Josephson junctions linking the qubits 
to the superconducting busbars, 
$D$ and $L$ are distances between the neighbouring qubits and between the two superconducting banks, respectively.
%,while $\hbar$ and $e$ are the Plank constant and the electron charge. 
It is convenient to use dimensionless variables: the inter-qubit distance $l=L/\lambda$,  vector potential 
$a_{z,n} = \pi D A_{z,n} /\Phi_0$,   time $\tau = \omega_J t$ and   energy $E_n = \mathcal{E}_n/E_J$, where the Josephson plasma frequency $\omega_J= 2 e I_c/(\hbar C) $ and the effective spatial scale $\lambda = c/\omega_J$. 
Then the energy per unit is 
\begin{eqnarray}
E_n = \Bigl( \frac{d \phi_n}{d\tau} \Bigr)^2 + \Bigl( \frac{d a_{z,n}}{d\tau} \Bigr)^2 - 2 \cos{\phi_n} \cos{a_{z,n}} 
\nonumber \\
+ \beta^2 \bigl( \frac{a_{z,n+1} - a_{z,n}}{l} \bigr)^2,
\end{eqnarray}

The quantization is performed by replacing the phases $\phi_n$ and their conjugate canonical momenta with operators \cite{zagoskin} and yields the Schr\"odinger equation 
\begin{equation}
i \hbar \frac{d}{dt} | \Psi_n^s \rangle = \mathcal{H}_{n}^{qubit} | \Psi_n^s \rangle = (\mathcal{H}_{n,int}^{qubit} + \mathcal{H}_{n,0}^{qubit}) | \Psi_n^s \rangle
\end{equation} 
with the qubit Hamiltonian $\mathcal{H}_{n,0}^{qubit}$ and field-qubit interaction $ \mathcal{H}_{n,int}^{qubit} $:
\begin{equation}
\mathcal{H}_{n,0}^{qubit} = - \frac{4 e^4 E_J}{\hbar^2 C^2 \omega_J^2} \frac{\partial^2}{\partial \phi_n^2}- 2E_J \cos{\phi_n},
\end{equation} 
\begin{equation}
\mathcal{H}_{n,int}^{qubit} = 2E_J \cos{\phi_n} ( 1- \cos{a_{z,n}})
\end{equation} 
We restrict our analysis to the two first energy levels of each island (which is justified for a realistic choice of charge qubit parameters). We will also neglect entanglement between qubits and seek the qubits' wave function in the {\em factorized} form: 
$ | \Psi_n^s \rangle =C_0^n(t) e^{-\frac{i \mathcal{E}_0 t}{\hbar}} |0 \rangle + C_1^n(t) e^{-\frac{i \mathcal{E}_1 t}{\hbar}} |1 \rangle$. Then
the Schr\"odinger equation for qubits reduces to
\begin{eqnarray}
i \frac{d}{d\tau} C_0^n(\tau)&=&s (1- \cos{a_{z,n}}) C_1^n(\tau)e^{-i s \epsilon \tau}, \nonumber \\
i \frac{d}{d\tau} C_1^n(\tau)&=&s (1- \cos{a_{z,n}}) C_0^n(\tau) e^{i s \epsilon \tau} .
\label{qub1}
\end{eqnarray} 
where $s =E_J/\hbar \omega_J$, $\epsilon$ is the dimensionless excitation energy (energy difference between the first and the ground levels \cite{zagoskin}).

Equations for the electromagnetic field in the transmission line can be derived by taking the expectation value of the total  energy (\ref{total}) in the  quantum state of the qubit subsystem, resulting in
the effective electromagnetic-field Hamiltonian $ \langle \mathcal{E}\rangle^{(em)}$ 
\begin{eqnarray}
\langle \mathcal{E}\rangle^{(EM)} = E_J \sum_n\Bigl( \langle \Psi_n^s | H_{n,int}^{qubit} | \Psi_n^s \rangle +\nonumber \\ \bigl( \frac{d a_{z,n} }{d \tau} \bigr)^2 + \beta^2 \bigl(\frac{a_{z,n+1} -a_{z,n}}{l} \bigr)^2 \Bigr) 
\end{eqnarray} 
where the coupling between qubits and fields is described by the term: 
\begin{eqnarray}
\sum_n \langle \Psi_n^s |H_{n,int}^{qubit} | \Psi_n^s \rangle = \sum_n (1- \cos{a_{z,n}} ) V_n(\tau) \label{int} \\
V_n(\tau) = C_0^{n \ast}(\tau) C_1^{n}(\tau) e^{-i s \epsilon \tau} + C_0^{n}(\tau) C_1^{n \ast}(\tau) e^{i s \epsilon \tau}. \nonumber
\end{eqnarray}
By using Hamilton's equations for the dimensionless vector potential $\partial/\partial\tau(2\partial a_{z,n}/\partial\tau)=-\partial \langle \mathcal{E}\rangle^{(EM)}/\partial a_{n,z}$, we derive
the effective sine-Gordon equation for the  line:
\begin{equation}
\frac{d^2 a_{z,n}}{d \tau^2} - \beta^2 \frac{a_{z,n+1} + a_{z,n-1} - 2a_{z,n}}{l^2} + V_n \sin{a_{z,n}}+\gamma\frac{da_{z,n}}{dt}=0 \label{sg}
\end{equation}
where we have added the phenomenological damping term $\gamma da_{z,n}/dt$ (which can be introduced in the quantum Routh formalism\cite{zagoskin}).
This equation is quite similar to the standard discrete sine-Gordon equation (see, e.g., in Ref. \cite{discrjj}) for a classical 
Josephson transmission line, but with the key difference
that the ``effective critical Josephson current'', $V_n$, now depends on the quantum states of the charge qubits. 

In contrast to the perturbative analysis of Eqs.(\ref{qub1},\ref{int},\ref{sg}) in \cite{mi}, here we perform a numerical analysis in the case of strong coupling where new light-matter states can be found.

{\it Results.---\/} First we consider which state the system settles in at zero magnetic field $H_{\rm ext}=0$, imposing boundary conditions $H_{\rm ext}=(a_{z,1}-a_{z,0})/l=(a_{z,N}-a_{z,N-1})/l=0$. Here $N$ is the total number of qubits in the system. In order to enable the system to evolve to its stationary state we add a very weak noise. To our surprise the vacuum state with $C_{1}^{n}=0,\ C_0^n=1, a_{z,n}=0$ is stable only for relatively 
weak qubit-field couplings. Then the vacuum state becomes unstable, and the system evolves to the quasi-superradiant state with one or several solitons (depending on the system size and interactions) spontaneously arising. Even though the energy of the qubit system itself and the energy of the 
magnetic field both increase (Fig. 2c), the total energy of this QS state decreases due to interactions between the field and the qubits. Figure 2a shows the magnetic field and vector potential distribution in the QS state. Since the vector potential changes from $-\pi$ to $\pi$ along the quantum transmission line, we conclude that a magnetic field soliton carries one flux quantum similarly to a usual Josephson vortex in a standard long Josephson junction (see, e.g., \cite{tinkham}). However, the QS state has a more complex structure than a Josephson vortex, since the soliton-like fields' distribution is here accompanied with a spatial modulation of qubit state occupation probabilities $C_0^n$ and $C_1^n$ (Fig. 2b). Here the macroscopic, {\em classical} magnetic soliton depends on the {\em quantum} state of the qubits.  In contrast to the standard superradiant state
of two-level atoms in a cavity interacting with one or few modes, the soliton is an essentially nonlinear magnetic field distribution arising via a complex interaction of a very large number of field-modes.

In zero magnetic field the transition to the QS state breaks the system's symmetry by spontaneously choosing the soliton polarity 
(positive or negative). The external magnetic field eliminates the spontaneous degeneracy of the QS phase, and the transition point
from the Meissner state with no solitons to the soliton state is shifted: the critical coupling for the transition becomes 
a function of $H_{ext}$). Figure 3 shows the 
evolution of the metamaterial phases when varying the external magnetic field. 
We consider the case when the vacuum state ($a_{z,n}=0$) is stable at $H_{\rm ext}=0$, but it
becomes unstable for quite a weak $H_{\rm ext}$ resulting in the formation of a one-soliton state (Fig. 3, main panel - the remagnetization loop 
and magnetic field distribution (A)). With a further increase of the external magnetic field $H_{\rm ext}$ 
 a sequence of the structural transitions to the phases with larger numbers of solitons occur 
accompanied with jumps in the trapped magnetic flux $\Phi$. When the external field $H_{\rm ext}$ decreases 
from a certain maximal value, other sequence of the structural transitions occurs first
decreasing the number of the ``positive-polarity'' solitons followed 
by formation and increasing the number of the ``anti-solitons'' or solitons with the opposite magnetic field direction. 

It is worth noting that only states with an odd numbers of solitons appear. More importantly, at zero external field $H_{\rm ext}=0$, the system still keeps one soliton on the steady remagnetization curve (external (blue) loop in Fig. 3) indicating that the formally stable initial Meissner state with zero 
solitons in the system is actually a metastable state. Thus, the vacuum state in this case
is metastable in contrast to the QS ground state (which can be called the ``dressed-vacuum'' state) 
for this value of the field-qubit coupling. Therefore, the transition from the vacuum to the
QS phase in one-dimensional quantum metanaterials is the first rather than the second order phase transition 
in contrast to the superradiant transition for the system of two level ``natural'' atoms and one-mode electromagnetic field in a cavity \cite{superrad}. 

Finally we consider the dependence of the QS transition on the sample size (the number of qubits in the quantum Josephson transmission line). Using the
same set of parameters as in Fig. 3, we simulate the steady remagnetization curve $\Phi(H_{\rm ext})$ for a shorter chain (see Fig. 4). As one can see, the state
at zero external field is always the vacuum (or the Meissner phase) with no solitons in the sample. Only at a high enough external magnetic field $H_{\rm ext}$, 
the soliton state (here with a single soliton) can be formed, but it becomes unstable with decreasing $H_{\rm ext}$ (on the returning branch of the 
remagnetization curve) before $H_{\rm ext}$ drops to zero. 

Remarkably, the unusual ``butterfly-like'' loops similar to those obtained here appear in several apparently different systems, including crossing vortex lattices \cite{nm2}, magnetic vortices in nano-discs \cite{novosad} and thermal atomic switches \cite{memrist}. For all such systems the ``butterfly-like'' loops originate
either due to a nontrivial interplay of fluctuations with driving or due to the complex nature of vortex pinning in the bulk and on surface. 
In our particular case, the quantum metamaterials provide a ``quantum pinning'' for Josephson-like vortices on the inhomogeneity 
formed by fluctuating qubit occupation numbers of ground and excited states (which are mobile and can diffuse through the transmission line).

{\it Conclusions.---\/} We predict a new state of matter for quantum metamaterials --- the quasi-superradiant soliton phase--- when the coupling between 
electromagnetic fields and qubits crosses a threshold (the critical coupling), which can be tuned by the external fields resulting in a series of structural
superrradiance transitions and a variety of remagnetization loops. The seeming violation of the "no-go" theorem for charge-based circuit QED can be attributed to the more complex structure of the field modes in the system. Note that the behaviour of flux-qubit based quantum metamaterials investigated so far is qualitatively similar to that predicted for the charge-based ones. Given that such metamaterial prototypes have already been fabricated, though on a small scale \cite{science-lasing,macha}, and the "no-go" theorem \cite{no-go-general} does not apply to them, it will be worthwile to investigate the QS transition in such etructures as well. Further investigation of this new state can hopefully give a new perspective on a long standing puzzle regarding the superradiant phase of two level atoms interacting with several modes of electromagnetic fields in a cavity. 

SES acknowledge support from Leverhulme Trust. AMZ was supported in part by the EPSRC grant EP/M006581/1 and by the Ministry of Education and Science of the Russian Federation in the framework of Increase Competitiveness Program of NUST«MISiS»(No. K2-2014-015).

\end{document}